\documentclass[papersize,journal]{IEEEtran}
\usepackage{amsmath,amssymb}
\usepackage{algorithmic}
\usepackage{algorithm}
\usepackage{array}
\usepackage{textcomp}
\usepackage{stfloats}
\usepackage{url}
\usepackage{verbatim}
\usepackage{graphicx}
\usepackage{cite}
\usepackage{multirow}
\usepackage{xcolor}
\usepackage{subfigure}
\usepackage{soul}
\begin{document}

\title{Improving One-Shot Transmission in NR Sidelink Resource Allocation for V2X Communication}

\author{Hojeong Lee and Hyogon Kim}

\maketitle
\thispagestyle{empty}

\begin{abstract}
The Society of Automotive Engineers (SAE) has specified a wireless channel congestion control algorithm for cellular vehicle-to-everything (C-V2X) communication in J3161/1. A notable aspect of J3161/1 standard is that it addresses persistent packet collisions between neighboring vehicles. Although the chances are slim, the persistent collisions can cause so called the wireless blind spot once the event takes place. Then the involved vehicles cannot inform their presence to neighboring vehicles, an extremely dangerous condition for driving safety. J3161's solution to the problem is stochastic one-shot transmission, where the transmission occasionally occurs in a resource that is not originally reserved. Through the one-shot transmission, the worst-case packet inter-reception time (PIR) is bounded and the wireless blind spot problem can be effectively mitigated. Interestingly, the standard one-shot transmission scheme does not resolve the persistent collision relation itself. This paper shows that by breaking out of the relation as soon as the persistent collision condition is identified, vehicles can improve the worst-case PIR by approximately 500 ms, the number of packet collisions per persistent collision event by 10\%, and the number of total collisions by 15\% to 57\% over the standard one-shot transmission.
\end{abstract}


\section{Introduction}
\IEEEPARstart{T}{he} 3rd Generation Partnership Project (3GPP) has been standardizing Cellular V2X (C-V2X) communication
technology since Release 14. One of the core features of the C-V2X standards is the sidelink communication that enables direct communication between vehicles. The wireless resource for the sidelink communication can be allocated in either centralized or distributed manner. The base station orchestrates the allocation in the former (called Mode 1), whereas vehicles autonomously allocate the resource using a distributed algorithm in the latter (called Mode 2) \cite{38214}. Because vehicles could be out of network coverage, Mode 2 is considered the base mode. The distributed resource allocation algorithm used in Mode 2 is called the Sensing-Based Semi-Persistent Scheduling (SB-SPS) \cite{38214}. 

As the name implies, the SB-SPS algorithm (henceforth ``SPS” for convenience) utilizes the sensed resource use pattern of neighbor vehicles to avoid the resource locations where the neighbors are expected to transmit. It also utilizes each neighbor’s explicit resource reservation information for its next transmission written in the control part of the packet \cite{38212}. Once a host vehicle decides on a resource least likely to be used by its neighbors, it uses the same resource for a certain number of subsequent packet transmissions without additional signaling, hence the name Semi-Persistent Scheduling. In this paper, we will call this series of packets using the same frequency resource over the average one-second period by the name of a ``packet run.” Note that the same resource can be kept in the next packet run with the resource keep probability $0 \le P_k \le 0.8$ \cite{38214}, so it can be a few seconds before a vehicle re-selects a different resource \cite{blindspot}.

\begin{figure}[htbp]
\centering
\includegraphics[width=0.9\linewidth]{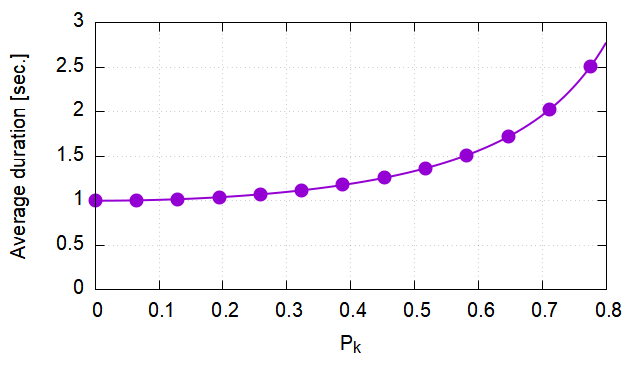}
\caption{Average time until breakout from repeated packet collisions relation}
\label{fig:avgcollen}
\end{figure}

An issue with SPS is that it cannot completely eliminate packet collisions. For instance, two vehicles that came to re-select a resource almost simultaneously can pick the same resource. If their message transmission periods are the same, the packet collision will repeat at every subsequent message transmission. The higher the resource keep probability $P_k$, the longer the repeated packet collisions. At a given resource keep probability $P_k$, the number of packet runs that retain the same frequency resource at a vehicle is geometrically distributed. Then, the two vehicles will break out of the consecutive packet collisions relation as the shorter of the two packet runs expires and the vehicle with the shorter run re-selects. Here, the average number of the shorter packet run is given by
\begin{equation}
\mathbb{E}[\min(l_1,l_2)] = \frac{1}{2p-p^2} \label{eq:breakout}
\end{equation}
for $p = 1-P_k$. According to Irwin-Hall distribution, the number of packet transmissions in the packet runs tends to a normal distribution centered at $l\overline{X}$ where $X$ is the run length of each packet run. For instance, $\overline{X}=10$ if the messaging period is 100 ms. As Fig. \ref{fig:avgcollen} shows, the average duration in Eq. (\ref{eq:breakout}) can easily persist for over a few seconds once the repeated collision event takes place, especially as $P_k \rightarrow 0.8$. The worst-case duration can be longer. Note that blindly reducing $P_k$ is not a solution, either, because it makes re-selections more frequent in SPS. Consequently, it causes more packet collisions and generally leads to poor packet delivery performance \cite{jeon18}.

The cost of a long, repeated-collision event is potentially high. The event renders the involved vehicles to be unrecognized by neighboring vehicles and by each other during the persistent collision event. Onboard sensors, such as radars, LiDARs, and cameras may still be able to maintain awareness in many situations, but in others like the non-light-of-sight (NLoS) positions, neighboring vehicle movements become difficult to trace under the consecutive packet loss event. Therefore, in safety-critical situations, this communication lull may push a vehicle into dangerous driving conditions \cite{blindspot}, which might even be worse off than not relying on the C-V2X safety communication at all.

In order to address the persistent packet collision problem, the Society of Automotive Engineers (SAE) J3161/1 stipulates stochastic One-Shot Transmission mechanism for C-V2X communication environment \cite{j3161}. Under the scheme, vehicles occasionally transmit their packets in a resource that is not originally reserved (see Fig. \ref{fig:one_shot}(a)). Through the one-shot transmission, the worst-case packet inter-reception time (PIR) is bounded and the wireless blind spot problem can be effectively mitigated. Interestingly, however, the standard one-shot transmission scheme does not resolve the persistent collision relation itself. Such passive approach neglects unnecessary packet collisions thereby degrading packet delivery performance under SPS. 

In this paper, we demonstrate that by exiting the relation as soon as the persistent collision condition is identified, the vehicles can further improve PIR and packet reception ratio (PRR) than in the standard one-shot transmission. Compared with the duration of repeated collisions in SPS (Fig. \ref{fig:avgcollen}), the proposed scheme reduces the worst-case PIR to less than 3 seconds at the farthest required communication range of 320 meters \cite{37885} for $P_k=0.8$. Note that this performance is approximately the minimum duration in SPS. Therefore, the proposed scheme significantly improves the long-tail PIR values. Compared with the J3161/1 One-Shot mechanism, the proposed solution reduces the worst-case PIR by approximately 500 ms, the number of packet collisions per persistent collision event by 10\%, and the number of total collisions by 15\% to 57\% over the standard one-shot transmission. 

The rest of the paper is organized as follows. Section \ref{sec:related} briefly discusses the prior work on resolving the problem of repeated packet collisions. In particular, it includes the previous efforts to improve on the performance of one-shot transmission. Section \ref{sec:sol} provides the background on SPS resource allocation and the standard J3161/1 one-shot transmission. It then discusses the proposed modification to SPS. Section \ref{sec:eval} evaluates the performance of SPS, one-shot, and the proposed scheme in terms of PRR, PIR, and the collision statistics. Finally, Section \ref{sec:conclusion} concludes the paper.

\section{Related work}\label{sec:related}
Bazzi et al. \cite{blindspot} proposed to curtail semi-persistent resource use at a fixed length, overriding repeated reselection of the same resource over a certain limit, even when it is prescribed by a high probability of keeping the same frequency resource. As a side-effect, any persistent collisions are limited to a deterministic length. But the proposed limit is significantly long than that stipulated in the one-shot transmission in J3161/1. 
Gholmieh et al. \cite{gholmieh2021c} analyzed that continuous collisions of packets can be reduced without noticeable loss of packet reception ratio (PRR) performance through one-shot transmission. In Fouda et al. \cite{fouda2021interleaved}, the authors analyzed the performance according to the range of one-shot counters. It showed that the performance of the tail of information age (IA) and inter-packet gap (IPG) was better when the one-shot counter range was [2..6], which is smaller than [5..15]. Saifudd et al. \cite{saifuddin2023addressing} showed that under J3161/1 congestion control, a timer-based one-shot transmission performs better than the standard counter-based scheme. This is because  when congestion control of J3161 intervenes, Inter-Transmit Time (ITT) increases. This implies that as the vehicle density increases, the time it takes for the one-shot counter to reach zero also increases. Consequently, the duration required to escape from consecutive collisions becomes longer, particularly compared to situations with lower vehicle density. 

Unfortunately, none of the counter-based and timer-based one-shot transmission methods proposed so far do not resolve the persistent collision condition itself. However, vehicles can ascertain whether other vehicles are utilizing the same resource through one-shot transmission. Therefore, it is neither desirable nor necessary to resume transmission on the resource identified as involved in the persistent collision condition. The enhancement discussed below that by breaking out as soon as the persistent collision condition is identified, packet delivery performance is indeed improved.

\thispagestyle{empty}
\section{Solution approach}\label{sec:sol}
\subsection{Background}
Under SPS, each vehicle runs a re-selection counter that is decremented upon each transmission. Once the frequency resource for the next packet run is re-selected, the counter is randomly initialized to a value in 
\begin{equation*}
\bigg[5\times \frac{100}{\max(RRP,20)} :\ 15\times \frac{100}{\max(RRP,20)} \bigg]
\end{equation*}
where $RRP$ is the resource reservation period in SPS \cite{38321}. In this paper, we will assume $RRP=100$ ms for illustrative purpose because it is the most popular messaging period for safety-related V2X applications, although the proposed scheme also applies to other RRP configurations.

SAE J3161/1 also defines a congestion control algorithm, where the gap between the transmitted packets called the Inter-Transmission Time (ITT) is controlled as follows \cite{j3161}:
\begin{equation}\label{eq:TxR_1}
ITT\  {\rm [s]} = \begin{cases}
0.1, & \overline{VD}\le25\\
\overline{VD}/250, & 25<\overline{VD}<150\\
0.6, &150\le \overline{VD}
\end{cases} 
\end{equation}
where the average vehicle density $\overline{VD}$ is obtained from the instantaneous $VD$ as 
$\overline{VD(t)} \leftarrow 0.05 \times VD + 0.95 \times \overline{VD(t-1)}$. $VD$ is the number of vehicles within 100 m of the observing vehicle.

\subsection{One-shot transmission in J3161/1}
The stochastic one-shot transmission in J3161/1 effectively mitigated the continuous collision problem by providing a mechanism to occasionally step aside and use a resource different from the currently reserved one. It is executed more frequently than re-selection, and is executed regardless of whether or not a persistent collision event is ongoing. In case the persistent collision event is underway, the one-shot transmission fragments the long series of packet collisions into many smaller ones. Compared with the unmodified SPS, therefore, neighbor vehicles can occasionally have a chance to notice a formerly unheard vehicle due to consecutive collisions.

According to SAE J3161/1, each vehicle additionally runs an one-shot counter $C_O$. It is randomly initialized to a value in $[2:6]$, and decremented with each transmission under SPS. Let $R_{sps}$ denote the resource allocated by SPS, and $R_{os}$, by one-shot. Based on the values of $C_R$ and $C_O$, the following logic is executed:
\begin{enumerate}
\item If $C_O=0$ and $C_R = 0$
    \begin{itemize}
        \item Transmit in $R_{os}$ instead of in $R_{sps}$; Initialize $C_O$
        \item Perform re-selection conditioned on $P_{keep}$; Initialize $C_R$ 
    \end{itemize}
\item If $C_O=0$ and $C_R \ne 0$
    \begin{itemize}
        \item Transmit in $R_{os}$ instead of in $R_{sps}$; Initialize $C_O$
        \item Do not decrement $C_R$
    \end{itemize}
\item If $C_O \ne 0$ and $C_R = 0$
    \begin{itemize}
        \item Perform re-selection conditioned on $P_{keep}$
        \item If moving, initialize $C_R$ and $C_O$ 
        \item If not moving, initialize only $C_R$ 
    \end{itemize}
\end{enumerate}
In case neither counter is 0, normal SPS-reserved transmission takes place and both counters are decremented by 1. Notice that there is no provision to exit the persistent packet collision condition itself.

\thispagestyle{empty}
\subsection{Proposed enhancement}
We propose an enhancement that curtails the persistent collision event as soon as an involved vehicle finds it. The vehicle then performs the SPS re-selection, exiting the collision relation. In this paper, we assume that the one-shot resource $R_{os}$ is randomly selected among the candidate resources recommended by SPS when $R_{sps}$ was selected, and that is within 3 slots after $R_{sps}$. This is to minimize the interference to other vehicles' SPS transmissions that could happen at the chosen one-shot resource. Also, limiting the time offset of the one-shot transmission to within 3 slots of the originally reserved resource is to minimize the impact on the packet delay budget (PDB). If no resource exists that is recommended by SPS within the time limit, random selection is performed with the same time constraint. Depending on applications, the PDB constraint may be less strict. Then, an one-shot resource can be selected with larger offsets, reducing the potential of packet collisions. Since the proposed scheme utilizes the information already available after $R_{sps}$ selection, it does not incur additional processing overhead that would be incurred if we separately invoked the complex SPS processing. In the enhancement, in Condition (2) in the logic presented above, we replace
\begin{itemize}
        \item \textbf{Do not decrement $C_R$}
\end{itemize}
with
\begin{itemize}
        \item \textbf{If other vehicle is observed to transmit in $R_{sps}$, perform SPS re-selection; Initialize $C_R$ and $C_O$}
        \item \textbf{Otherwise, do not decrement $C_R$}
\end{itemize}

\begin{figure}[htbp]
\centering
\subfigure[One-shot Transmissions according to J3161/1]{\includegraphics[width=\linewidth,height=0.5\linewidth]{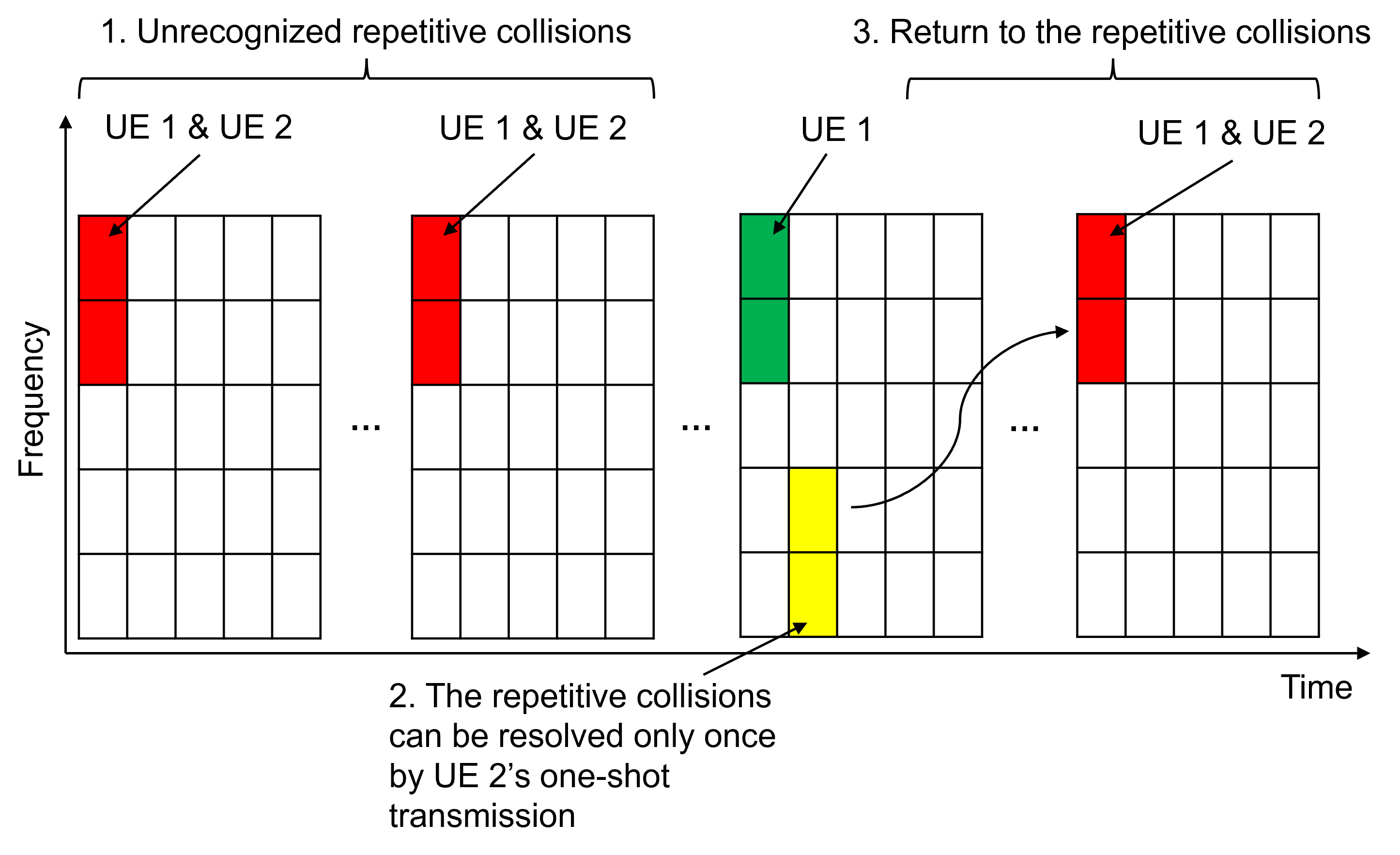}}
\subfigure[Proposed method using One-shot Transmissions]{\includegraphics[width=\linewidth,height=0.5\linewidth]{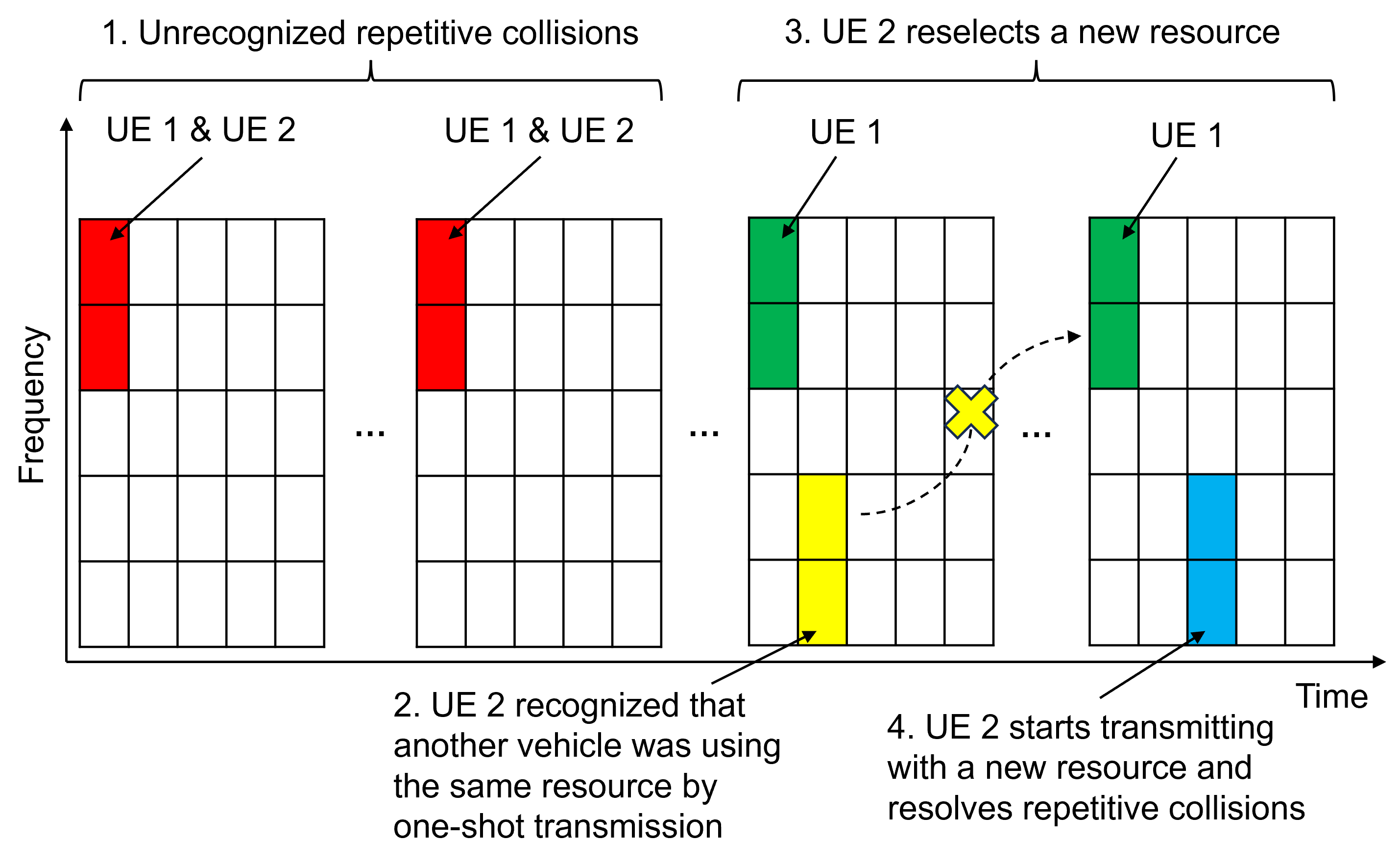}}
\caption{Standard one-shot transmission scheme vs. proposed enhancement}
\label{fig:one_shot}
\end{figure}

Fig. \ref{fig:one_shot}(a) and (b) illustrate the difference between J3161/1's one-shot transmission and the enhancement. In the figure, `UE' refers to a user equipment, which is the on-board V2X communication unit on a vehicle. The standard one-shot scheme in Fig. \ref{fig:one_shot}(a) detects the persistent collision condition by intermittently performing one-shot transmission. However, even after it finds the condition through the one-shot transmission, the involved vehicles UE1 and UE2 do not exit the condition but keep using the collision resource $R_{sps}$ for future transmissions until the next re-selection. In the proposed enhancement (Fig. \ref{fig:one_shot}(b)), however, the persistent collision condition itself is resolved by the vehicle that first realizes the condition and immediately moves to the one-shot resource $R_{os}$. Namely, UE1 first performs the one-shot transmission and realizes that its previous resource $R_{sps}$ has another user UE2, so UE1 performs a re-reselection through SPS and moves to a different resource. The persistent packet collision condition is thus broken.

Finally, when three or more vehicles are involved in a persistent packet collision condition, the condition can be found by the one-shot performing vehicle through high Received Signal Strength Indicator (RSSI) value even if the transmission(s) there is not decoded. But the probability of such multi-vehicle persistent collisions is deemed low, so we focus on the two-vehicle case in this paper.

\section{Performance evaluation}\label{sec:eval}
In this section, we evaluate the impact of the proposed enhancement in terms of packet delivery performance. To this end, we conducted simulation experiments comparing three schemes in the C-V2X environment, as follows:
\begin{enumerate}
\item SPS without one-shot transmission (``SPS'')
\item One-shot transmission as defined by J3161/1 (``One-shot'')
\item Proposed solution (``Proposed'')
\end{enumerate}
As to the metrics, we utilized the distribution of Packet Inter-Reception time (PIR), number of consecutive collisions and collision events, and Packet Reception Ratio (PRR). PIR and PRR are the most significant measures of packet delivery performance in V2X communication and are defined as follows \cite{37885}:
\begin{itemize}
    \item PRR: ratio of successfully received packets to the total packets transmitted within 320 meters and binned into 20-meter segments 
    \item PIR: time elapsed between two successive receptions of two different packets transmitted from the same vehicle within the same range of 320 meters 
\end{itemize}

\thispagestyle{empty}
\subsection{Simulation setting}
For simulation experiments, we employed LTEV2Vsim \cite{ltev2vsim}, an open-source simulator designed for C-V2X environment. 
Specific simulation parameter values are summarized in Table \ref{tab:params}. Note that the presented results are obtained with J3161/1 congestion control enabled whereas hybrid automatic repeat and request (HARQ) disabled because we consider periodic broadcast traffic such as Basic Safety Message (BSM) \cite{J2735}. Also, we do not use the blind retransmission because it doubles the bandwidth use, which can cause a severe congestion and the J3161/1 congestion control will increase the ITT. It will make information update much less frequent, betraying the very purpose of periodic beaconing through BSM packets.

\begin{table}[hbpt]
\caption{Simulation parameter configuration\label{tab:params}}
\centering
%\scriptsize
\begin{tabular}{|l|l|l|}\hline
& Parameter &	Value \\ \hline
\multirow{12}{*}{PHY}
    & Bandwidth (MHz) & 20 \cite{j3161}\\
    & No. subchannels & 10 \\
    & Subchannels/transport block & 2 \\
    & Antenna gain (dB) & 3  \\
    & Tx power (dBm) & 20 \cite{j3161} \\
    & Noise figure of receiver (dB) & 9 \\
    & Pathloss model & WINNER+B1 \cite{37885}\\
    & Shadowing distribution & Log-normal\\
    & Shadowing std. dev. (dB) & 0 (LOS), 4 (NLOS)\\
    & Antenna height (m) & 1.5 \\
    & MCS index & 7 \\ \hline
\multirow{6}{*}{MAC} & re-selection counter value & [5..15] \\
    & One-shot counter value & [2..6]  \\
    & Selection window width (ms) &  100 \\
    & Congestion control &  J3161/1 \\
    & HARQ &  Off \\
    & Resource keep probability  & 0.8 \cite{38214} \\ \hline
\multirow{1}{*}{Application}
    & Message size (bytes) & 250 \\      \hline
\multirow{3}{*}{Vehicles}    & Traffic density (veh./km) & [100..500]\\ 
    & Speed of vehicles (km/h) & $\mu$ = 50, $\sigma$ = 3 \\ 
    & Road length (km) & 2 \\ \hline
\end{tabular}
\end{table}

\subsection{Packet inter-reception time}

\begin{figure}[htbp]
\centering
\subfigure[$\rho=100$]{\includegraphics[width=\linewidth,trim={0 0 0 0},clip]{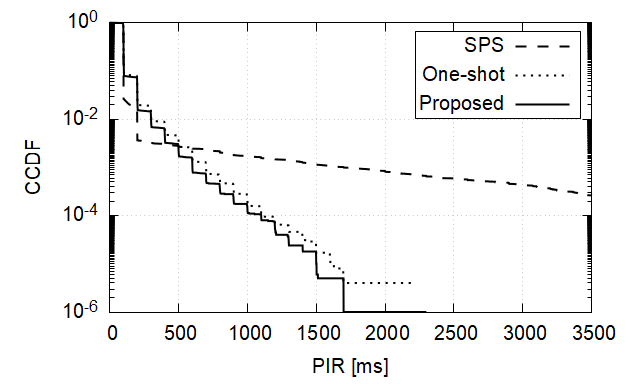}}
\subfigure[$\rho=200$]{\includegraphics[width=\linewidth,trim={0 0 0 0},clip]{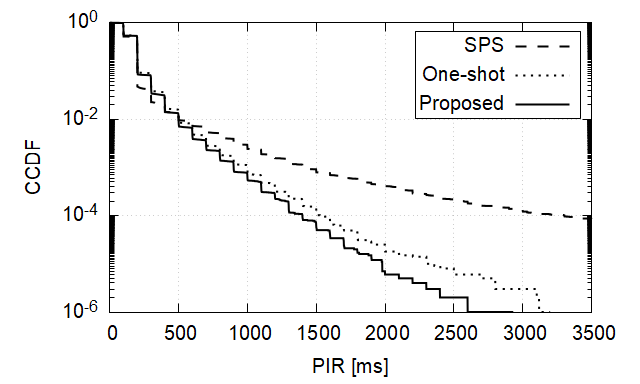}}
\subfigure[$\rho=300$]{\includegraphics[width=\linewidth,trim={0 0 0 0},clip]{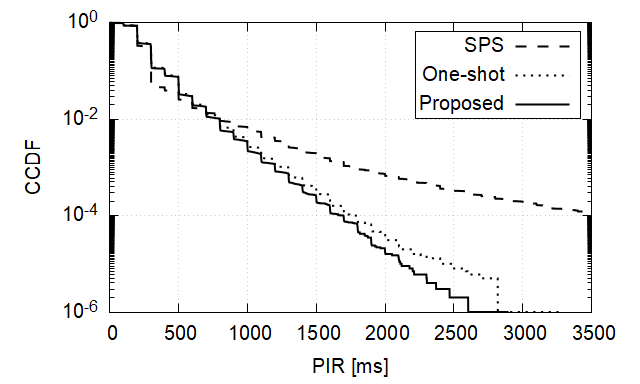}}
\subfigure[$\rho=400$]{\includegraphics[width=\linewidth,trim={0 0 0 0},clip]{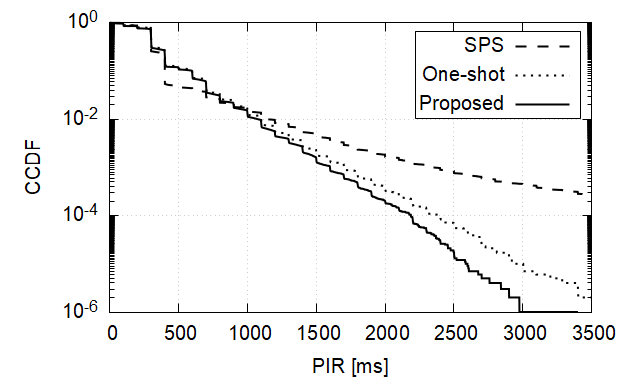}}
\caption{PIR distribution for different vehicle density environments}
\label{fig:pir}
\end{figure}

Fig. \ref{fig:pir} shows how One-shot and Proposed improve the tail PIR distribution over the unmodified SPS. For $\rho=100$ veh./km (or equivalently $VD=20$ in Eq. (\ref{eq:TxR_1})), One-shot and Proposed can respectively limit the PIR below 2,200 and 1,700 ms at $10^{-6}$ probability whereas SPS has high PIR with slowly decreasing probability. In fact, SPS cannot suppress the PIR under 3,500 ms even at $10^{-4}$ probability in any traffic density. For other vehicle densities, we can observe similarly significant improvement enabled by the two one-shot schemes. However, the improvement is not free. In particular, the standard one-shot transmission trades extremely large PIR values for more numerous intermediate PIR values by breaking long consecutive collision events into multiple medium-length events. For example, with $\rho=100$ veh./km, PIR values between 200 ms and 400 ms have increased in number. Note that it is an inevitable cost to avoid the dire situations where the periodic beacons from two close vehicles suffer from effectively becoming ``ghosts" for an extended period of time \cite{blindspot}. Comparing One-shot and Proposed, the latter always outperforms former. This is because Proposed eliminates the residual consecutive collision events after the first one-shot opportunity.

A further improvement could be made possible by forcing a one-shot transmission as soon as the resource re-selection is performed. Assuming that consecutive packet collisions between more than two vehicles are rare, we can force each vehicle perform a one-shot transmission for the first packet transmission after a re-selection with a probability of 1/2. In case there is indeed another vehicle that selected the same resource with the same RRP, one of the vehicles can identify the consecutive collision condition and move to other resource with probability 1/2. However, it may have a complicated affect on SPS, so we leave it for a future work.

\subsection{Collision statistics}

\begin{figure}[htbp]
\centering
\subfigure[
Average no. of packet collisions per persistent collision event]{\includegraphics[width=\linewidth,trim={0 0 0 0},clip]{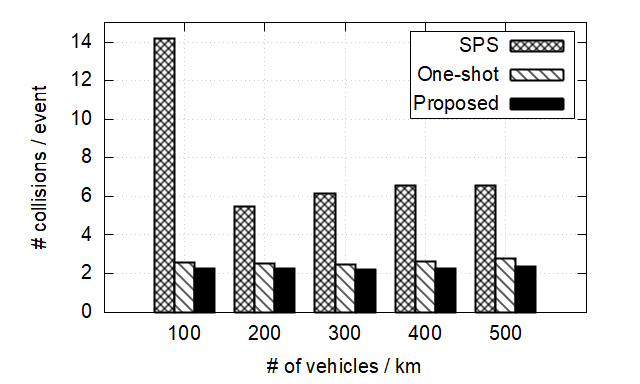}}

\subfigure[Total number of packet collisions observed throughout simulation]{\includegraphics[width=\linewidth,trim={0 0 0 0},clip]{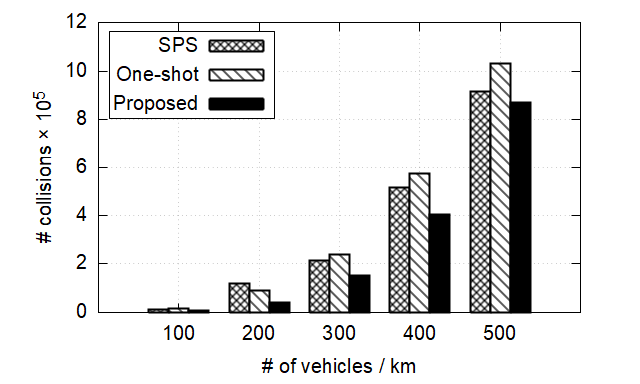}}
\caption{Packet collision statistics}
\label{fig:num}
\end{figure}

Fig. \ref{fig:num} compares the collision characteristics of the three schemes. Fig. \ref{fig:num}(a) shows the average number of consecutive collisions in a persistent collision event. Although the tail distribution information is not available as in Fig. \ref{fig:pir}, the average run length can be shown to decrease with the proposed enhancement. Although Proposed significantly reduces the number of such events over One-shot, the run length per event only slightly differs between them.

We can easily estimate the run length for Proposed. It is the minimum $X$ of two (colliding) one-shot runs $X1,X2 \sim U[2..6]$, i.e., 
\begin{equation*}
\mathbb{E}[X]=\mathbb{E}(\min[X1,X2]) = \sum_{i=2}^5 2\cdot \frac{1}{5} \cdot \frac{6-i}{5} = 2.4
\end{equation*}
where a one-shot run is defined by the run length of packet transmissions before the one-shot counter ($C_O$) expires. 
In case of One-shot, it is slightly higher than Proposed. It can be illustrated by an example. Suppose $C_O(u)=2$ and $C_O(v)=6$ for two vehicles $u$ and $v$. With Proposed, the number of collisions would be 2. However, with the standard one-shot, the number of collisions before the first one-shot by $u$ is 2, but the next ones could be longer than this minimum length. For instance, if $u$ chooses $C_O(u) > 2$, the next collision run will have a length larger than 2, increasing the average.

For $\rho = 200$ veh./km, both One-shot and Proposed reduce the average run length of packet collisions compared to SPS, which is expected. But an unexpected result is that SPS has smaller number of consecutive collisions than at $\rho=100$ veh./km. This is because the J2945/1 congestion control begins to kick in, changing the packet collision dynamics. The table below shows that the average ITT is approximately 155 ms for $\rho=200$ veh./km. 
\begin{center}
\footnotesize
\begin{tabular}{|c||c|c|c|c|c|}\hline
$\rho$ (veh./km) & 100 &	200	& 300	& 400	&500 \\ \hline
Average ITT (s) & 0.101	& 0.155	& 0.232	& 0.313	& 0.387 \\ \hline
\end{tabular}
\end{center}

\thispagestyle{empty}
Because only discrete RRP values $k\cdot 100\ (k\ge 1)$ ms are allowed over 100 ms in SPS \cite{38214}, vehicles are transmitting at an ITT of either 100 ms or 200 ms. When the vehicles with different RRPs meet with persistent collisions, one with 200 ms RRP will still suffer the consecutive collisions as before whereas the one with 100 ms RRP will experience a collision every other transmission. The latter will not qualify as consecutive collisions. However, for those vehicles with 200 ms RRP, the collision run length will be also shortened because the counterparts with 100 ms RRP decrement their $C_R$ twice as fast. It cuts off the consecutive collisions faster than the situation of the both have the same RC value reduction. As $\rho$ increases further, the population ratio between the vehicles with different RRP values affect how fast a persistent collision event is cutoff by the small RRP counterpart.

Fig. \ref{fig:num}(b) reveals an important aspect of one-shot transmission schemes. In particular, One-shot
increases the total number of packet collisions compared to SPS. This is natural because one-shot transmissions occur outside the resource originally reserved ($R_{sps}$). When the persistent packet collision does occur, each one-shot eliminates a collision. But with far higher probability the event does not occur, and the one-shot transmissions may only collide with SPS transmissions from other vehicles. Indeed, the figure shows that the latter outweighs the former. This is the second cost aspect of the one-shot transmission, from the perspective of packet collisions. Unlike the standard one-shot, however, the proposed enhancement has even lower number of packet collisions than SPS. This is because once an one-shot transmission that identifies, the rest of the consecutive packet collisions retained by both SPS and the standard one-shot scheme are avoided.

\subsection{Packet reception ratio}
\begin{figure}[htbp]
\centering
\includegraphics[width=\linewidth,trim={0 0 0 0},clip]{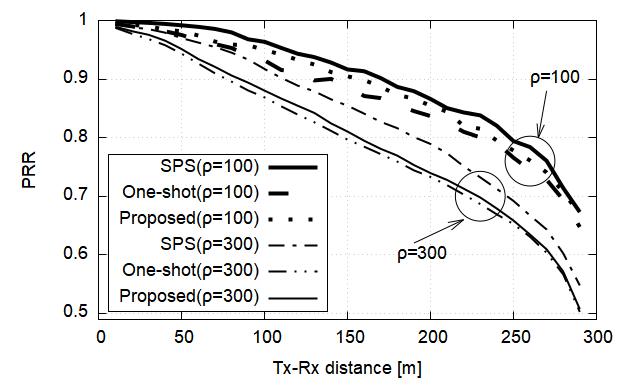}
\caption{PRR distribution for different vehicle density environments}
\label{fig:prr}
\end{figure}

The main purpose of one-shot transmission is to avoid the rare but dangerous V2X communication outages at all costs. However, it is worthwhile to consider how adversely average packet delivery performance is affected by it. Fig. \ref{fig:prr} compares the PRR for the three schemes. For readability, we show for only $\rho=100$ and $\rho=300$ veh./km. Other vehicle densities show the same qualitative result. Although Gholmieh et al. \cite{gholmieh2021c} argued that PRR performance is hardly affected, there is small PRR drop according to our simulation results. We believe that it is unavoidable because one-shot transmission frequently leaves the reserved resource wasted and instead uses a non-reserved resource. Because the proposed enhancement reduces such incursions through early re-selection, it leads to slightly better PRR over the original one-shot transmission scheme. In essence, the proposed scheme does not degrade the PRR performance compared to the standard one-shot transmission scheme.

\thispagestyle{empty}
\section{Conclusion}\label{sec:conclusion}
In safety-critical driving situations, preventing rare but dangerous communication outages is also important. The SAE J3161/1 standard prescribes so called the one-shot transmission that stochastically allows vehicles to step aside from potential persistent packet collisions and reveal their presence. However, it is too passive because it does not go as far as using the standard re-selection mechanism readily provided by the Semi-Persistent Scheduling (SPS) upon finding the persistent collision condition. This paper proposes an enhancement to the one-shot scheme to perform the re-selection. Proposed enhancement not only improves the PIR tail distribution but reduces the number of packet collisions over both SPS and the one-shot scheme. Moreover, it leads to slightly better PRR performance than the standard one-shot scheme.

\section*{Acknowledgments}
This research was supported by the MSIT (Ministry of Science and ICT), Korea, under the ICT Creative Consilience program (IITP-2023-2020-0-01819) supervised by the IITP (Institute for Information \& communications Technology Planning \& Evaluation.

\vfill

\end{document}